\documentclass[runningheads,a4paper]{llncs}

\usepackage{amssymb}
\usepackage{graphicx}

\usepackage{url}
\urldef{\mailsa}\path|{nitriant, pfrangos} @central.ntua.gr|
\urldef{\mailsb}\path|petros@math.ntua.gr|   
\newcommand{\keywords}[1]{\par\addvspace\baselineskip
\noindent\keywordname\enspace\ignorespaces#1}

\begin{document}

\newtheorem{def1}{Definition}
\newtheorem{exam}{Example}
\newtheorem{prop1}{Proposition}

\mainmatter  

\title{OTS2DBC: An attempt to combine Design by Contract with Behavioral Algebraic Specifications\footnote{This paper was supported by the THALIS project "Algebraic Modeling of Topological and Computational Structures and Applications". The Project "THALIS" is implemented under the Operational Project "Education and Life Long Learning" and is co-funded by the European Union (European Social Fund) and National Resources (ESPA). }}

\titlerunning{OTS2DBC}

%
%
\author{Nikolaos Triantafyllou\and Petros Stefaneas \and Panayiotis Frangos}
\authorrunning{OTS2DBC}


\institute{National Technical University of Athens, \\
Iroon Polytexneiou 9, 15780 Zografou, Athens, Greece\\
\mailsa\\
\mailsb\\
\url{}
}

%
%

\toctitle{OTS2DBC}

\tocauthor{OTS2DBC}

\maketitle

\setcounter{secnumdepth}{5}
\setcounter{tocdepth}{3}

\begin{abstract}
Abstract. Design by Constract (DBC) has influenced the development of formal specification languages that allow the mix of specification and implementation code, like Eiffel, the Java Modeling Language (JML) and Spec$\#$. Meanwhile algebraic specification languages have been developing independently and offer full support for specification and verification of design for large and complex systems in a mathematical rigorous way. However there is no guarantee that the final implementation will comply to the specification. In this paper we proposed the use of the latter for the specification and verification of the systems design and then by presenting a translation between the two, the use of the former to ensure that the implementation respects the specification and thus enjoy the verified properties. 

\keywords{CafeOBJ, JML, Observational Transition Systems, Design by Contract,  Behavioral Specifications, }

\end{abstract}

\section{Introduction}

Over the last decade we are seeing an increased research interest towards formal Behavioral Interface Specification Languages (BISs), implementing Design by Contract (DBC), for languages such as Java and C$\#$.  What is even more encouraging is that major companies like Google and Microsoft are involved in the development of such languages (cofoja[28], spec$\#$[30]). The basic concept in such languages is a contract. A contract specifies the obligations and requirements of methods and method callers. Contracts are defined in the programming language itself and are translated into executable code by the compiler. Some of these languages also provide a wide range of tools that implement either real-time assertion checks (RAC) or static verification tools (SV).  Meanwhile the development of specification languages independent of a particular implementing language is continuing. Over the decades many such languages and theorem provers have been developed. They adopt much sounder mathematical bases and provide powerful theorem provers that are able to verify the most complex of properties. In addition they can use various levels of abstraction and specify and verify design. For example in the family of algebraic specification languages the research and applications of languages like OBJ, BOBJ, CafeOBJ, Maude etc. has been substantial.  
One could argue that either approach is better. From a programmers point of view it is preferable to have a specification language that is close to the programming language, that is easy to learn and can be used without having to learn difficult mathematical concepts.  On the other hand from a design engineer or mathematician point of view the soundness of the theorem provers, the ability to reason about design and not implementation and so on is more important. In this paper we attempt to combine the two approaches and hopefully adopt the strengths of both while overcoming some of their weaknesses.  In a nutshell we propose that the specifications are initially generated in an algebraic specification language capable of defining Observational Transitions Systems (OTS) specifications (like CafeOBJ). The desired system properties, be they safety or liveness, are proved with regard to this specification using the sound methods provided by this framework. Then this specification is translated into a DBC specification language, like the Java Modeling Language (JML). Finally the programmer can create the implementation based on the JML contracts and the JML compiler checks whether the implementation satisfies the specification either during run-time or statically. The choice of JML is due to purely historical reasons, since the features required  can be found in most modern DBC languages for Java such as modern Jass [29] etc. The OTS/CafeOBJ method was chosen because concepts like inheritance, the use of multiple objects/systems etc. are defined naturally here]. 

\subsection{Related Work}
Predating our work here, are [16, 17, 18 and 29]. In [16] the authors propose an automatic translation from CafeOBJ specifications to Java template code (which contains only the signatures of methods the class needs etc.). In [17] they go a step further to generate automated code for the behavioral part of the specification. But as they state because an OTS/CafeOBJ specification is a state based specification, the output of this automated translation may not be a suitable statement for the method we are trying to implement in Java. In [18] the authors provide design notations of an OTS specification implemented in CafeOBJ (Meta-OTS/CafeOBJ) and an OTS implemented in Java (Meta-Java/CafeOBJ), then they provide a specification of the translation rules from Meta-OTS/CafeOBJ to Meta-Java/CafeOBJ in CafeOBJ. This allows them to finally verify the correctness of the translation.  The work in [29] is the closest to ours. The authors see B-machines as the most suitable language for the specification of Java programs and present an automated translation from B-machines to JML. This approach is very close to what we propose here, the main difference is that we believe OTS/CafeOBJ specifications are better suited for the specification of Object-orientated systems due to their natural support for inheritance, the use of multiple objects and so on, features that are not supported at the moment in [29]. 

\subsection{Contribution}
The contribution of this work as far as CafeOBJ is concerned lies in the fact that using the proposed methodology we can create an implementation (in Java) of an OTS/CafeOBJ specification that satisfies it. CafeOBJ is independent from the choice of a particular implementation language. This is both a strength, generality of the method and so on, but at the same time is a weakness because it provides no guarantee that the implementer will conform to the specification so that properties that were proved will still hold by the implementation. By defining a formal translation to JML we can guarantee the production of Java code that conforms to the CafeOBJ specification, and consequently holds the same properties (there are tools to check that the Java code written by the programmer satisfies the JML specification).  We believe this is better than a direct translation to the implementing language. This way the designer can handle the design issues without getting into unnecessary implementation/optimization details that might not be easy to express in a specification language. 

One could argue why not only use JML then or some other DBC language for Java? An important downside of JML is the lack of support for the current version of Java. Other DBC for Java languages like Cofoja and modern Jass support Java 7, however they lack the verification tools. Also, while Java has proven its power, no language is perfect. So it might be beneficial to use different languages for different parts of the implementation. By only using JML it is not possible to reason about parts of the system that are not implemented in Java. By specifying in a more general and abstract language such as OTS/CafeOBJ we can reason about the design of the whole system and its properties and then implement the components in the appropriate language such as Java or C$\#$. Of course there should be such translations from OTS/CafeOBJ to say spec$\#$ in order to use C$\#$, but these translations are left for the future. 

Secondary to the above, our approach provides a way to minimize specification. Only a small subset of the implementation has to be specified, the methods/classes that correspond to the OTS/CafeOBJ specification. For the rest of the implementation the only restriction is that the extra methods/objects have to be called/used from within contracted methods. In addition the proofs in the OTS/CafeOBJ method are computer human interactive. As a result when a proof fails it conveys information as to why that happened, this can result in the redesign to the system and a deeper understanding of the problem at hand. Closing, a key feature of this approach is the reusability of proofs as well as specifications. Objects are combined in ways that preserve their properties, so it will be easy to create libraries of verified code, that can be used freely and reduce the time required to specify new systems.

The rest of this paper is organized as follows. Section 2 gives an overview of related work. Section 3 introduces the reader to Design by Contract (DBC) and JML, section 4 presents some key elements of Observational Transition System (OTS), CafeOBJ and how OTSs are expressed with CafeOBJ terms. Section 5 presents our mapping from OTS to JML/Java. Section 6 the translation from OTS/CafeOBJ to JML/Java and the final section concludes the paper with some future goals.

\section{Short Introduction to JML}
Design by contract (DBC) is a method for developing object orientated software first proposed in 1992 in [4]. The corner stone of this approach is that the class's methods and the clients (programs) that invoke them have a so call contract between them. On the one hand the invoker of a method guarantees certain pre-conditions before calling the method and on the other hand the method is able to guarantee that certain post-conditions will hold after its execution. The use of such pre and post conditions to specify the behavior of programs dates back to Hoare's paper [5] in 1969. The basic idea of this Hoare logic is that of a triplet {P} C {Q} where P and Q are assertions, (the pre and post-conditions) and C is a command of the program. The main contribution of DBC is that these contracts are executable, meaning that they are defined in the program code in the programming language and are translated into executable code by the compiler.  This makes it possible for any violation of the contracts that occur while the program is running to be detected immediately. 
JML is a formal Behavioral Interface Specification Language for Java [1] that can be used as a design by contract (DBC) tool for Java programs. A JML specification is created using special annotated comments, which start with the sign @. The specification of the obligations of the client  is denoted by using the key words //@ requires and the methods post-conditions, that the implementer has to meet, are denoted using  //@ensures. A contract in JML is a set of pre and post conditions for a method, such contracts can be used to as an abstraction of a methods behavior. As in Eiffel [6], JML uses Java's expression syntax to write the predicates that are used to express the assertions and pre and post conditions. The lack of expressiveness of the Java language to write formal specifications is solved by extending Java expressions with the required specification contracts, for example quantifiers [1]. In addition JML is designed to be used with a range of tools [7, 8]. The range of these tools is from DBC support to runtime assertion checking and to verification of specifications using theorem provers [1]. 
The semantics of a JML method's specification is that when the method is invoked in a state where its pre-conditions are met, then one of the following two things will happen. Either the method terminates normally, in which case the post-conditions of the specification are guaranteed, or there method throws an exception, in which case the exception thrown must be permitted by the specification (either by default or explicitly) and the exceptional post-conditions  must be satisfied (these conditions are denoted by the signals clause) [1]. 
Invariants in JML are used as a way to reduce the size of a specification. An invariant is a JML property that must hold in all visible states that a method is invoked. Here a visible state is defined as the parts of the code where the method is invoked. Consequently these invariants must hold at the end of each constructor execution and at the beginning and end of all method invocations. The main difference between invariants and pre, post-conditions is that invariants apply to all subtypes through specification inheritance, while predicates that appear in all pre, post-conditions are not inherited as part of the specification of any new methods that may be added through subtyping [2]. 
JML supports several kinds of quantifiers in assertions:  universal quantifier ($\diagdown$forall), existential quantifier ($\diagdown$exists), generalized quantifiers ($\diagdown$sum, $\diagdown$product, $\diagdown$min and $\diagdown$max) and a numeric quantifier ($\diagdown$$num\_of$) among others. Referring to the value of a field before the invocation of the method is done using the old clause ($\diagdown$old). Extra constructs such as implies (==$>$), follows from ($<$==), if and only if (<==>) are defined as welll [2]. 
Some of the more advanced notions for behavioral specification using JML include method Purity and Frame Properties, Datagroups, Model Fields and Ghost Fields. Due to the restriction of DBC, for having only side effect free methods as parts of the assertions of the contracts, only query methods can be used. This is defined in JML as method Purity and is declared using the keyword pure as an annotation before the name of the method [3]. By Frame properties we mean that any variables not mentioned in the specifications post-conditions should remain unchanged after the invocation of the method. In JML the assign clause in the contract specifies what parts of the system may change by this method [3]. Frame properties are restrictive in that since they cannot be extended through specification subtyping. To solve this JML uses datagroups. A datagroup is an abstract piece of an objects state that may be extended by future subclasses by simply stating what fields belong to it.  To connect such an abstract field with a field of the implementation we use the JML represents clause. A datagroup is assigned to every model field so that we can use them in the assignable clauses. If we wish to provide additional state, which may or may not be related to the existing state we use a Ghost field. A ghost field is a specification only field that unlike the model field can be assigned a value inside the specification, throw the set construct [2].

\begin{table}[t]
\centering
\begin{tabular}{p{12cm}}
\hline	
public class Person $\lbrace$ \\
private /*@ non$\_$null @* String name; \\
private int weight; \\
/*@ public invariant !name.equals("") @ $\&\&$ weight $>$= 0; @*/ \\
\\
//@ ensures $\diagdown$result != null; \\
public String getName() $\lbrace$return name; $\rbrace$ \\

//@ ensures $\diagdown$result == weight; \\
public /*@ pure @*/ int getWeight() $\lbrace$retrun weight; $\rbrace$ \\

/*@ requires kgs $>$= 0; \\
   @ requires weight + kgs $>$= 0; \\
   @ ensures weight == $\diagdown$old(weight + kgs); \\
   @*/ \\
public void addKgs(int kgs) $\lbrace$ weight  = weight + kgs; $\rbrace$ \\

/*@ requires n != null $\&\&$ !n.equals(""); \\
   @ ensures n.equals(name) \\
  @ $\&\&$ weight == 0; @*/ \\
public Person(String n) $\lbrace$ \\
name = n; $\rbrace \rbrace$ \\
\hline
\end{tabular}
\caption{JML/Java Class example}
\end{table}

The JML compiler (jmlc) is an extension to the Java compiler that compiles Java programs annotated with JML specifications into Java byte-code [1]. The compiler includes in the byte code run time assertion checking instructions. Using these instructions, it is possible to check the assertions of the specifications such as pre, post-conditions on run time [1]. This feature falls in the category of Runtime Assertion (RAC) tools. We must keep in mind however that the specification may be incorrect, but the code implementation consistent to it. Note that the assertion checker is transparent when no inconsistences occur. As stated in [2] one of the major advantages of the jmlc, over other runtime assertion tools like Eiffel and Jass, is that it supports abstract specifications written in JML in terms of model fields, ghost fields, and model methods. Any run time violation is reported by some error (this is the Eiffel approach [1]).  But the tool support is not limited to RAC. Static verifications (SV) tools such as ESC/JAVA2 [3], are available that allow the verification that no violations of the verification will occur during run time using logical techniques. As analyzed in [3] these tools preform what is called extended static checking [10, 11], compile time checking that goes well beyond type checking. In particular ESC/JAVA2 supports the full of JML and is capable of checking for inconstancies between the code and the JML annotations. ESC/JAVA2 is neither sound nor complete deliberately and uses a non- interactive theorem prover. Another similar tool is LOOP which can be used to verify JML annotated code [12],  as the authors of [3] state it can be very labor intensive but it allows the verification of more complicated properties that cannot be check automatically by ESC/JAVA2. For similar intent tools such as JACK we refer you to [3].
Some of the more interesting properties in real life systems are the so called Safety and Liveness properties. A safety property states that something bad will never happen, while a liveness property states that something good eventually will happen. The former means that the system will never reach a state that a desired predicate does not hold, while the later that the system eventually will reach a state where the desired predicate holds. These types of properties are not easy to either specify or verify in JML. The author of [13] originally propose an extension of JML that allows the verification of such temporal properties called JTPL (Java Temporal Pattern Language), then through the use of a tool called JAG (JML Annotation Generator) translate this back into JML annotations ensuring the correctness of the translation [14].
Closing this section, we present a simple Java class annotated with JML, to give a first feel of the language. Assume we wish to specify that a class defining a person and that each such person object is required to have a valid name and a valid weight, also we desire that the new objects have a weight of zero. We also wish for this class to have methods that returns the weight of the person, his name and a method that allows us to add weight to him. Such a Java Class annotated with JML can be seen in table1.

\section{Behavioral Specifications and Observational Transition Systems}

Hidden algebra is an approach for giving semantics particularly for concurrent distributed object orientated systems, as well as for software engineering in general [15]. This approach is based on equations rather than other formalisms, like higher order logic for example. The reason behind this decision is that the proofs supported by this calculus allow maximal simplicity of mechanization and at the same time this approach allows for adequate expressiveness [15]. Hidden algebra is a generalization of process algebra and transition systems that by including non-monadic operations, allows the use of equations that contain methods or attributes that are parameterized by data [15]. The key of this approach is the concept of behavioral satisfaction. This means that we focus on how systems behave in response to a given set of experiments and not on the implementation details of those systems. So here the state space of an object (system) is regarded as a kind of black box, where we can only observe the state of it by using these experiments, or attributes or observation operators as they are called. This is a version of behavioral type in the object oriented paradigm, but like the authors of [15] we will refer to it as behavioral specification.  
In the heart of this approach are the behavioral objects, that are used to model a system and we will informally present here. As the author of [19] states, a behavioral object is simply a special case of behavioral specifications that formally specifies the state space of the system together with two set of operators on the state space. The set of actions (or methods) that are capable of changing the state of the system and the set of observations (or attributes) that are the experiments mentioned before, operators who take as arguments instances of the state space and return data types, the result of the experiments. Informally two states of such a behavioral object are considered equal when all observers return the same values in those states. Also two behavioral objects are equivalent when there is an isomorphism between the implementations modulo, the same state space, the same actions, and the same behavioral equivalence between the states [19]. Behavioral objects defined in this manner allow for the definition of several types of composition operators on behavioral objects [19]. For extensive definitions of hidden algebra, behavioral specification, and behavioral object compositions we refer the interested reader to [15, 19] as they are outside of the scope of this paper.

An Observation Transition System (OTS) is a transition system that can be written in terms of equations and is a proper sub-class of behavioral specifications. Assuming there exist a universal state space Y, and that each data type we wish to use is already defined in terms of initial algebra an OTS S is defined as the triplet $S=<O,I,T>$, where [20]:

\begin{itemize}

\item	O is a finite set of observers. Each $o \in O$ is a function $o: Y \rightarrow D$, where D is a data type and may differ from observer to observer. Given an OTS S and two states $u_1,u_2 \in Y$ the equivalence between them wrt. S is defined as $\forall o \in O, o(u_1) =_S o(u_2 )$. 

\item 	I: is the set of initial states for the system, i.e. $I \subseteq Y$.

\item 	T: is finite set of transition functions. Each $\tau \in Ô$ is a function $\tau : Y \rightarrow Y$, such that $\tau(u_1 )= \tau(u_2 )$ for each $[u]\in Y/_{=_S}$ and $u_1,u_2 \in [u]$. $\tau(u)$ is called the successor state of u with respect to S. Also with each $\tau$, comes a condition $c-\tau$, called the effective condition of $\tau$, such that $\tau(u) =_S u$ if $\neg c-\tau(u)$. 
\end{itemize}

Observers and transitions are usually parameterized, generally they are denoted as $o_{i_1 \dots i_m }$  and $\tau_{j_1 \dots j_n}$ provided that there exist data types $D_k$ and $k \in \lbrace i_1, \dots i_m, j_1, \dots j_n \rbrace$ with  $m,n \geq 0$.

\subsection{CafeOBJ in a nutshell}

CafeOBJ is a formal specification language that is able to handle the verification of algebraic specifications through various techniques. These include the computer human interactive verification of safety properties [23], liveness properties [24] and a framework for fully automated verification [25], finally support is provided for finding counter examples [26].  These techniques have been heavily applied for the verification of a plethora of real life systems in many different fields. 
CafeOBJ is based on multiple logical foundations but mainly on initial and hidden algebras [21]. Static aspects of systems are specified in terms of initial algebras and dynamic aspects are specified in terms of hidden algebras. CafeOBJ incorporates observational (behavioral) specifications. 
The basic building blocks of all CafeOBJ specifications are modules.  The keyword \texttt{mod*}, denotes the beginning of a module, with loose semantics (many possible models) and mod! if we wish to denote a module with tight semantics. Each module defines a sort (either visible or hidden), the name of the sort defined is declared by enclosing it between \texttt{[}and \texttt{]} in the case of visible sorts, and between \texttt{*[}and \texttt{]*} in the case of hidden sorts. The definition of a module starts with its signature, the names of the sorts it defines, an ordering on them, and the set of operators on the sorts. In addition a module defines a theory on that signature, i.e. a set of equations constructed using the elements of the signature. An underscore $\_$ in the definition of the operators indicates the place where an argument is put, for example \texttt{$\_$=$\_$: Sort1 Sort1 -$>$ Bool.} declares that the equality operator will take two arguments and return a Boolean value (Bool is a built in module/sort for the CafeOBJ system) . Finally modules can be imported using the keywords \texttt{pr( MODULENAME)}.

An OTS is transferred to CafeOBJ in a natural way. Based on the hidden algebra approach, a visible sort denotes an abstract data type, while a hidden sort denotes the state space of an abstract machine. On hidden sorts two kinds of operators can be applied; action operators and observation operators. An action operator is used to change the state of the abstract machine while we can only use observation operators to observe the inside of the machine. Declarations of observation and action operators begin using the keywords \texttt{bop} or \texttt{bops}, for the other operators we use the keywords \texttt{op} and \texttt{ops}. We define the operators using equations. The start of an equation is denoted with the keyword \texttt{eq} and the start of a conditional equation with the keyword \texttt{ceq}. The machine of CafeOBJ regards these equations as left to right rewrite rules and uses them to reduce complex expressions to simpler ones. The universal state space of an OTS, Y, is denoted by a hidden sort, say \texttt{H}. An observer $o_{i_1 \dots i_m} \in O$ is denoted by a CafeOBJ observation operator. For the following we assume that there exist visible sorts $V_k$ and $V$ denoting the data-types $D_k$ and $D$ respectively with $k=i_1 \dots i_m$. The observation operator that corresponds to $o_{i_1 \dots i_m }$ is defined in CafeOBJ as: \texttt{bop o : H $V_{i_1} \dots V_{i_m} -> V$}.

Each member of the set of initial states of the OTS is denoted by a constant of the hidden sort, i.e., \texttt{ op init: -> H}. Now suppose that the value of the  observer $o_{i_1 \dots i_m}$  is $F(x_{i_1}, \dots,x_{i_m} )$ in the initial state \texttt{init}, where $x_{i_j}$  are variables of the visible sorts $V_{i_j}$ .  This will be expressed with the following equation:  \texttt{eq o(init,$x_{i_1},...,x_{i_m}$ ) = f($x_{i_1},…,x_{i_m} )$}, where $f(x_{i_1},…,x_{i_m} )$ is a CafeOBJ term denoting $F(x_{i_1},…,x_{i_m} )$. Each transition of an OTS $\tau_{j_1 \dots j_n} \in Ô$ is denoted by a CafeOBJ action operator. Once again assuming visible sorts the CafeOBJ action operator that corresponds to the transition is define as: \texttt{bop $\tau$ : H  $V_{j_1} \dots V_{j_1m}$ -> H}. A transition rule (action operator) will successfully change the values returned by the observers if it is applied to a state where its effective condition hold. This is written in CafeOBJ notation as: \texttt{ceq o($\tau(S,x_{j_1 },... x_{j_m} ),x_{i_1},...,x_{i_m} )$ = e-$\tau$(S,$x_{j_1}, \dots, x_{j_m}, x_{i_1}, \dots ,x_{i_m} )$  if c-$\tau(S,x_{j_1 }, \dots ,x_{j_m } )$}. where S is a CafeOBJ hidden sort variable denoting an arbitrary system state, $x_k$ are variables for the visible sorts $V_k$, also $\tau(S,x_{j_1},...,x_{j_m})$ denotes the application of transition rule $\tau_{j_1 \dots j_m }$ to state S, i.e. the successor state of S and \texttt{e-$\tau$(S,$x_{j_1},\dots ,x_{j_m},x_{i_1} \dots x_{i_m} )$} is a CafeOBJ term denoting the value returned by the observer $o_{i_1 \dots i_m}$ in this successor state. Finally $c-\tau(S,x_{j_1 }, \dots ,x_{j_m})$ is a CafeOBJ term denoting the effective condition of $\tau_{j_1 \dots j_m}$. Let us point out here that the transitions do not change the values of the observers when the effective conditions do not hold.

\section{OTS/CafeOBJ to Java with JML}
\subsection{OTS to Java Class}

The main constructs of a JML specification are JML \emph{invariants}, \emph{constraints} and finally \emph{requires} and \emph{ensures} clauses.  JML allows the declaration of specification only variables on the form of \emph{ghost variables} and \emph{model fields}. Finally it is possible in JML to declare what a method is permitted to affect using Frame properties and the \emph{assignable} clause. 
In this section we will use the above  to provide an implementation of an Observational Transition System (OTS) in JML. This translation will ensure that the properties proved for the OTS will still hold for the Java implementation. This is achieved because the Java implementation has to comply with the JML specification when using the jmlc compiler or can be statically verified using the various tools. In this subsection we will consider only the case of one OTS that will not contain other OTSs. 
For the following definition we assume that there exist predefined data-types in Java, either built-in or user defined, for the corresponding abstract data-types of the corresponding OTS. As an OTS can be regarded as a type of black box so does the corresponding generated Java object 'hide' from us its full state and allows us to characterize it only by the use of pure methods, that correspond to the observers.  

\begin{def1}

\end{def1} Given an OTS  $S=<O,I,T>$ we define an implementation of S in Java/JML as a \emph{Java Class S} such that:

\begin{itemize}

\item 	For each observer $o_i\in O$ we define a pure method $o_{p_i}$ such that if  $o : Y \times D_{o_1} \times \dots \times D_{o_m}  \rightarrow D_o$ then the signature of $o_{p_i}$ will be: \texttt{public /*@ pure @*/ $D_o \,  o_{p_i}  (D_{o_1} d_1, \dots, D_{o_m} d_m)$ } .

\item  	For each initial state $u_{init}  \in I$ , we define a Constructor method $Cu_{init}()$, such that if     $o(u_{init},y_1, \dots ,y_n )=y_{o_{init}}$ then $Cu_{init}()$ is JML annotated with no requires clause and in the post-condition for all observers we add clauses of the form \texttt{//@ ensures $o_{p_i} (y_1, \dots, y_n )$ == $y_{o_{init}}$  ;} 

\item In addition we define a \emph{Behavioral Equality} method. This is a method \texttt{public boolean equals(S another)},and is annotated with JML code that ensures that if all the observers for two objects are equal then they are considered equal. So for each observer method $o_{p_i}$ of the class S the following annotation is added:
\texttt{//@ esnsures} $\diagdown forall (D_{o_1} d_1 \dots D_{o_m} d_m ; \,  this.o_{p_i}(d_1, \dots, d_m) == another.o_{p_i}(d_1, \dots, d_m) )$ \texttt{=> $\diagdown$result == true ;}

\item Also we define a \emph{Deep Copy} constructor, with signature \texttt{public S(S another)}. This is annotated with a requires clause \texttt{//@ requires another!= null}, stating that the object to copy cannot be null. Also this constructor ensures that the observers of the two objects will be equal and that the object created and the object copied will not point to the same place in the memory. These are defined with the following JML code:  \texttt{//@ ensures}  $\diagdown forall (D_{o_1} d_1 \dots D_{o_m} d_m ;$  $this.o_{p_i}(d_1, \dots, d_m) == another.o_{p_i}(d_1, \dots, d_m)$  $\&\& \dots \&\&$ $this.o_{p_k}(d_1, \dots, d_m) == another.o_{p_k}(d_1, \dots, d_m))$ $\&\&$ $(this != another) ;$

\item 	For each $\tau_i\in T$, $\tau_i:Y D_{t_1}… D_{t_n}  \rightarrow Y$,  we define a method with signature: \texttt{public S $\tau_i (D_{t_1}  x_{t_1}, \dots ,D_{t_n}  x_{t_n} )$}. Also in the body of the method we have a local ghost variable to store the pre-state of the object \footnote{this is necessary because the $\diagdown old$ JML command only stores the pointer to the object in the pre-state and not the object it self. We believe that this technique is acceptable since the impact of object creation is often overestimated, and can be offset by methods like garbage collection, etc. [31].Finally note that the ghost field could be as easily been implemented as a regular variable/object, however we prefer to produce as little implementation code as possilbe here}. This is achieved using the deep copying  constructor we defined above and an annotation like \texttt {//@ set ghost S temp = new S(this) ; }. This method is also annotated with two JML contracts:

\begin{enumerate}
	
	\item 	For the first contract, in the "requires clause" we have a set of statements that define the conditions under which the effective condition of the transition rule c-$\tau_i$, holds (using the pure methods that correspond to the observers).  Also in this contracts ensures clause we define:
	\begin{enumerate}
		
		\item that the result of the method is the new state of the object, \texttt{//@ $\diagdown$ result == this ;}
		
		\item In the post-condition for each observer  $o_n \in O$ such that $o_n(\tau_i(u,y_{i_1}, \dots, y_{i_n} ),$ $y_1, \dots, y_k)= y_n $             the following is added \texttt{//@ensures  $o_{p_n} (y_1,\dots, y_k )$ == $y_n$ ;}. If for $y_n$ a reference to the             pre-state is necessary that is made using the \texttt{temp} ghost field.

	\end{enumerate}

	\item 	The second contract contains in its requires clause the statements necessary to  make $c-\tau_i$ false and an assignable clause that makes certain that the call of the method in a system state that does not satisfy the effective condition will not change the state of the system. This can be defined as\footnote{another way to model this would be as an expepsional post-condition.}: \texttt{//@ assignable $\diagdown nothing$;} 
\end{enumerate}

\end{itemize}

An example of this translation is given in table 2, where the specification in OTS/CafeOBJ and the corresponding translation to JML/Java of a simple bank account system is given. The OTS/CafeOBJ code was taken from [19].

\begin{table}[h!]
	\centering
		\begin{tabular}{p{6cm} p{6cm}}
		\hline
		OTS/CafeOBJ                        & JML/Java \\
		\hline

		\texttt{mod* ACCOUNT}$\lbrace$                   &  \texttt{public class Account}$\lbrace$ \\
		\texttt{pr(INT)}                                 & \texttt{//@ public ghost Account temp;} \\
		\\
		\texttt{*[Account]*}                             & \texttt{//@ ensures balance() == 0;} \\
		                                                 & \texttt{public Account()}$\lbrace \rbrace$     \\
		\texttt{-- initial states}                       &  \\     
		\texttt{op initAcc : -> Account}                 & \texttt{public  /*@ pure @*/ int  balance()} $\lbrace \rbrace$ \\
		\\
		\texttt{-- observers}                            & \texttt{/*@ requires this.balance() + x >= 0 	;}  \\
		\texttt{bop money : Account -> INT}              & \texttt{@ ensures ($\diagdown$result.balance() == temp.balance() + x) $\&\&$ $\diagdown$result = this ;} \\

	                                                   & \texttt{@ also}   \\
		\texttt{-- transitions}                          & \texttt{@ requires this.balance()+x < 0 ;} \\
		\texttt{bop add : Account INT -> Account }       & \texttt{@ ensures $\diagdown$result.balance() == temp.balance() $\&\&$ $\diagdown$result = this   ; @*/} \\
		\texttt{op c-add : Account INT -> BOOL}          & \texttt{public Account add(int x)} $\lbrace$  \\
		\texttt{var I : Int}                             & \texttt{//@ set temp = new Account(this) ; }$\rbrace$ \\           
		\texttt{var A : Account}                         & \\
		                                                 & \texttt{//@ ensures (this.balance() == ac2.balance()) ==> ($\diagdown$result == true);} \\
		 \texttt{eq c-add(A,I) = read(A)+I >= 0 .}        & \texttt{public boolean equals(Account ac2)} $\lbrace \rbrace$ \\
		 \texttt{eq read(init) = 0 .}       &  \\
		 \texttt{ceq read(add(I,A) = I + read(A) if}     &  \texttt{//@ requires another != null ;} \\
	 	 \texttt{c-add(A,I) .}$\rbrace$                  & \texttt{//@ ensures (this.balance() == another.balance()) $\&\&$ (this != another);} \\
	                                                & \texttt{public Account(Account another)}$\lbrace \rbrace$ \\
		
		\hline			
		\end{tabular}
		\caption{Simple bank account system in OTS/CafeOBJ and JML/Java}
\end{table}

\subsection{Multiple OTSs to Multiple Objects}

In real life applications it is highly unlikely that an object orientated program will require the use of only one object. Usually multiple objects will work together to produce a single application. This feature is already covered in the behavioral specification paradigm. In such techniques reusability is a key feature. In object orientated programming reusability of the source code is the goal while in object oriented specification, the ability to be able to reuse both the code and the proofs is desired. In their paper [19] the authors provide ways to reuse code and proofs. Their approach is compositional. The key technical construct that allows this reusability is a special type of observer, the projection. The formalism in [19] is done via Hidden Algebra [22]. Here for uniformity we will continue using the OTS/CafeOBJ formalism since the relation between OTSs and Hidden Algebra is well established and the interchange here does not cause problems. 

In order to define an object composed by a set of other objects, we must define for each composing object a projection operator to get their states from the state of the composed object. These operators are subject to several mathematical conditions which are formally defined in [19]. Their essence is as follows [19]:
\begin{enumerate}

\item all transitions of all components are represented by transitions at the level of the compound OTS via these projection operators

\item each observer of the compound object is related via the projection operations to an observation of some component and finally

\item each constant state of the compound object is projected to a constant state on each component in the compound objects.

\end{enumerate}

Note that the previous intuitive definition implies that we have a fixed number of composing objects. The first characterization of this kind of composition is concurrency. In order to define concurrency we need the notion of a method group. The authors of [21] define that two transitions of a composed object belong to the same method group when they are related to the same composing object. Concurrent connection is now defined as the case where  if we take two transitions, which belong to different method groups, the order of application of these does not change the reached state.  Synchronized concurrent connection now is defined as the case where we have partial concurrency between the composing objects. Synchronization happens when either the projected state of a composing object is depended on the state of another composing object or when the transitions of the composed object change simultaneously states of several composing objects. A first result of this is that behavioral equivalence in composed objects is a conjunction of all the behavioral equivalences of the composing objects [21].

Static systems are a special case of Dynamic systems. In the latter the configuration of the system may change through the execution of the transitions. This is covered by allowing identifiers to manage object creation and deletion, i.e. parameterizing the projections. In [21] the authors define that a dynamic object can be created or deleted in a composed object and its initialization is handled via object identifiers. These operators conform to the conditions given above as well but in addition a projection operator for a dynamic object has an object identifier in its arguments. A second result of the paper is a refinement of the first and states that for the case of dynamic systems the behavioral equivalence of a composed object is a conjunction of all the behavioral equivalences of the composing objects [21, 19].

In OTS/CafeOBJ notation a parametrized projection is defined as, $\pi : ID \, Y \rightarrow Y_n$, where Y is the state space of the composite object, $Y_n$ is that of the $Class_n$ composing objects and $ID$ is a set of identifiers for them. This is transfered to JML/Java as a parametrized pure method that returns  $Class_n$ objects. 

\begin{def1}[Projection Methods]
In order to retrieve the state of a $Class_n$ object we define a pure method, with signature \texttt{public /*@ pure @*/ $Class_n$ getCn (int i)$\lbrace \rbrace$} , that is annotated with the following JML contract\footnote{This ensures that when the objects returned by the projection exist, they are different objects and not just different references to the same object.}: 
\begin{verbatim}
/*@   public normal_behavior
	  @   requires (i >= 0) ;
	  @   ensures (\forall int j ; 
	  @   ( ((getCn(i) != null) && (i !=j)) ==> getCn(i) != getCn(j)) ) ;
	  @*/
\end{verbatim}
\end{def1}

Before continuing we have to update the definition of the deep copy constructor so that it correctly copies the composing objects. For each class of composing objects \texttt{Cn} we add to the contract of the deep copy constractor from definition 1 the following annotations:

\begin{verbatim}
@ ensures (\forall int i; (this.getCn(i)!= null) ==> 
@ this.getCn(i).equals(another.getCn(i)) &&  (this.getCn(i) 
  != another.getCn(i)) ;
\end{verbatim}

For the following we assume that all base level objects we will use contain the equality and deep copy constructor methods we specified in definition 1. Bullet two, of the informal definition states the each observer of the composite object is equal to the composition of projections and chains of actions, observations of  a component object(s). By denoting with $ch_n$ the chains and by $f$ the composition, for an observer $o$ of the composite object in OTS/CafeOBJ notation a composite object observer will be defined by an equation of the form: $o = (\pi_{i};ch_{n_1}, \dots, \pi_{k};ch_{n_k})f$.

\begin{def1}[Composite Object Observers]
The previous composite object observer is defined in JML/Java as a pure method o with a signature \texttt{public /*@ pure @*/ $D_o$ o($D_1 y_1, \dots D_n y_n$, int $id_1, \dots,$ int $id_k$)$\lbrace $ //@ set temp = new S(this);$\rbrace $} that is annotated with the following contract:
\begin{verbatim} 
\*@ requires this.getC1(id1) != null  
  @ &&... 
  @ && this.getCk(idk)!= null ;
  @ ensures \result == f(temp.getC1(id1).chn1(), ..., 
  @ temp.getCk(idk).chnk()); 
  @*/
\end{verbatim}

\end{def1}

Where \texttt{chni()} are Java terms that correspond to the chains $ch_{n_i}$. Note that we must take extra care to ensure that the transition calls inside these chains do not change the actual states of the composing objects. This is why it is necessery to use the ghost field \texttt{temp}.

Before we give the definition of the transitions in a composite OTS we must update the behavioral equality method (\texttt{equlas}) so that it takes into account the projection methods. So we define that for all projection methods $getCn$ the following term is added to the contract of the method:
\begin{verbatim}
//@ ensures  (\forall int i; this.getCn(i).equals(another.getCn(i)) ) ;
\end{verbatim}

Bullet three of the informal definition states that, for each transition of the composite object $\tau$ and each composing $Class_n$ object, $c_{n_i}$ there exists a transition of the compoment object $\tau_{n_i}$ such that for state u of the composite object $\tau(u);\pi_{n_i} = \pi_{n_i}(u);\tau_{n_i}$. This is expressed in JML/Java with the following definition.

\begin{def1}[Composite Object Transitions]
At the first line of the body of the method we store the pre-state of the object in the ghost field \texttt{temp}. For each object, $c_{n_i}$, who's state is change by the transition we add in the requires clause of the contract: \texttt{//@ requires this.getCn(i)!= null}. Finally, the ensures clause for each such object we add: \texttt{//@ ensures this.getCn(i).equals(temp.} \texttt{getCn(i).$\tau_{n_i}$(...);)}. 

\end{def1}

In the dynamic compositional paradigm, the dynamic compound objects might add new actions which do not correspond to actions of the components. More over if we wish to handle a variable number of component objects two transitions are madantory in order to define the insertion and deletion of objects. In most cases\footnote{allthough more parameters might be required depending on the situation, for simplicity we assume that only the hidden sort and the is required here} the signatures of these methods will be \texttt{add : Y ID $\rightarrow Y$} and \texttt{delete : Y ID $\rightarrow$ Y}.

\begin{def1}[Adding and deleting components] 
The composite object transition that adds a new $Class_n$ component to the configuration of the system is defined having a signature \texttt{public S add(int id)$\lbrace \rbrace$} and is contracted as:

\begin{verbatim}
/*@ requires this.getCn(id) == null; 	
@ ensures \result.getCn(id).equals(new Account() ) && 
@ (\forall int j ; (j != id) ==> \result.getCn(id) != \result.getCn(id)) &&
@ \result == this;
@*/
\end{verbatim}

Also composite object transition that removes a component object from the configuration of the system, has as signature : \texttt{public S del(int id) $\lbrace \rbrace$} and is contracted as:
 
\begin{verbatim}
/*@ requires (this.getCn(id)!= null) ;
@ ensures \result.getCn(id) == null && 
@ \result == this; ;
@*/
\end{verbatim}

\end{def1}

To clarify all these definitions the previous example of the bank account is expanded in Appendix A, to a bank system where accounts are created and deleted dynamically and the users can add are withdraw money from their accounts. First we give the code in OTS/CafeOBJ notation and then the corresponding JML/JAVA code\footnote{all examples were implemented with the open JML plugin for eclipse, http://sourceforge.net/projects/jmlspecs/}.

\subsection{Inheritance}

One of the most attractive features of object orientated programming is inheritance.  Observation Transitions Systems are a proper subclass of behavioral specifications. The semantics of behavioral specifications are those of hidden algebra [22, 15]. When it comes to defining inheritance in the OTS framework it is more convenient to talk in terms of hidden algebras rather than OTSs. The state space of an object is represented as a hidden sort. In hidden algebra there exist two kinds of sorts. Visible sorts that represent the data part of the specification and hidden sorts representing the states of the objects.  A hidden signature of an order sorted algebra is defined as $(S,\leq,\Sigma)$ [19], where S is a set of sorts and $\Sigma$ is the set of operators on S, also we consider $H \subset S$, , the hidden sorts. The visible sorts, V, are defined simply as $S\setminus H$. In the above $(S,\leq)$ is a partial order set. An operator is denoted as $\sigma: w \rightarrow s$, , where $w \in S^*$ and $s \in S$. We call the input of $\sigma$, w, the arity of the operator and the result, s, the co-arity.  Now given such a signature a model of it interprets each sort $s \in S$ as set $A_s$, each subsort relation $s<s'$ to an inclusion $A_s \subseteq A_{s^{'}}$ and finally each operator $\sigma \in \Sigma_{s_1,\dots,s_n,s}$  to a function $A_{\sigma} : A_{s_1} \times \dots \times A_{s_n} \rightarrow A_s$.  In hidden algebra, inheritance is modeled via subsort relations [15]. This means that the space of states of the inheriting object is included in the space of states of the inherited object: enabling the inherited methods to act on the states of the inheriting object. 
In OTS notation the previous definition would mean that if we wish to denote that $S=<O,I,T>$ with the state space Y, is inherited by the OTS $S'=<O',I',T'>$ with the state space Y',  then  we must define that $Y' \subseteq Y$.  Such a specification would be translated into Java/JML by simply stating that the class that corresponds to the OTS S' extends the class that corresponds to the OTS S.

\section{Conclusions}

We have presented a translation between the OTS/CafeOBJ specifications and the JML/Java specifications. The goal of this was to allow for the specification and verification of the design of systems in OTS/CafeOBJ and at the same time guarantee that the Java implementation complies with the specification using the tools of JML. This approach allows us to use in our opinion the best of both worlds and accomplishes to overcome some of the shortcomings. 
In the future we will build tools that automatically translate an OTS/CafeOBJ specification to a JML specification to facilitate this process. Also similar translations and tools must be defined for other DBC specification languages, so that the various components of a large system that are implemented in various programming languages will comply with the original specification in OTS/CafeOBJ and thus enjoy the desired verified properties.

\section{APENDIX A}

\begin{verbatim}
mod* ACCOUNT-SYSTEM {
pr(ACCOUNT)

*[ AccountSys ]*

op init-account-sys : -> AccountSys -- initial state
bop add : UId Nat AccountSys -> AccountSys -- method
bop del : UId AccountSys -> AccountSys -- method
bop deposit : UId Nat AccountSys -> AccountSys -- method
bop withdraw : UId Nat AccountSys -> AccountSys -- method
bop balance : UId AccountSys -> Nat -- attribute
bop account : UId AccountSys -> Account -- projection

vars U U' : UId
var A : AccountSys
var N : Nat

eq account(U, init-account-sys) = no-account .
ceq account(U, add(U', N, A)) = add(N, init-account(U))
if U == U' .
ceq account(U, add(U', N, A)) = account(U, A)
if U =/= U' .
ceq account(U, del(U', A)) = no-account
if U == U' .
ceq account(U, del(U', A)) = account(U, A)
if U =/= U' .
ceq account(U, deposit(U', N, A)) = add(N, account(U, A))
if U == U' .
ceq account(U, deposit(U', N, A)) = account(U, A)
if U =/= U' .
ceq account(U, withdraw(U', N, A)) = add(-(N), account(U, A))
if U == U' .
ceq account(U, withdraw(U', N, A)) = account(U, A)
if U =/= U' .
eq balance(U, A) = read(account(U, A)) .
}

\end{verbatim}

\begin{verbatim}
public class AccountSYS {
	
	private ArrayList<Account> accounts ;
	//@ public ghost AccountSYS preObj ;
	
	//@ensures (\forall int i; this.getAcc(i) == null) ;
	public AccountSYS(){ }
	
	// Deep Copy Constructor
	/*@ public normal_behavior 
	  @ requires another != null ;
	  @ ensures (\forall int i; (this.getAcc(i)!= null) ==> 
	  @ (this.getAcc(i).equals(another.getAcc(i)) &&
	  @  (this.getAcc(i) != another.getAcc(i))  
	  @ && (this.getBalance(i) == another.getBalance(i))) ) ;
 	*/
	public AccountSYS(AccountSYS another){ }
	
	// 1. Projection OPERATOR
	
	/*@   public normal_behavior
	  @   requires (i >= 0) ;
	  @   ensures (\forall int j ; 
	  @   ( ((getAcc(i) != null) && (i !=j)) ==> getAcc(i) != getAcc(j)) ) ;
	  @*/
	public /*@ pure @*/ Account getAcc(int i) { }

	// 3.EVERY OBSERVER IS REPRESENTED BY AN OBSERVER AT THE COMPOSITE OBJECT
	
	//@ requires this.getAcc(id) != null ;
	//@ ensures  \result == this.getAcc(id).balance() ;
	public /*@ pure @*/ int getBalance(int id) {
	//@ set preObj = new  AccountSYS(this) ;
	}
	
	//2. ALL TRANSITIONS ARE REPRESENTED BY TRANSITIONS IN THE COMPOSITE LEVEL
	
	//@ requires (this.getAcc(id) != null) && (n >= 0) ;
	/*@ ensures  \result.getAcc(id).equals(preObj.getAcc(id).add(n)) 
	  @ && \result == this;
	  @*/
  public  AccountSYS deposit(int n, int id){
		//@ set preObj = new  AccountSYS(this); }
	
	//@ requires (this.getAcc(id) != null) && (n >= 0) ;
  //@ ensures  \result.getAcc(id).equals(preObj.getAcc(id).add(-n)) 
    @ && \result == this; ;
    @*/
  public  AccountSYS withdraw(int n, int id){
    	//@ set preObj = new  AccountSYS(this) ; 	}
	
	
	// Extra Composite Level Transitions
	//@ requires this.getAcc(id) == null; 	
	/*@ ensures \result.getAcc(id).equals(new Account() ) && 
	@ (\forall int j ; (j != id) ==> \result.getAcc(id) != \result.getAcc(id)) 
	@ && \result == this;
	@*/
	public AccountSYS add(int id){ 
	//@ set preObj = new AccountSYS(this); 	}
	

	//@ requires (this.getAcc(id)!= null) ;
	//@ ensures \result.getAcc(id) == null && \result == this;
	public AccountSYS del(int id) { 
		//@ set preObj = new  AccountSYS(this) ; 	}
}


\end{verbatim}

\end{document}